\documentclass[11pt]{article}
\def\withcolors{0}
\def\withnotes{1}
\def\withindex{0}

\usepackage[T1]{fontenc}
\usepackage[utf8]{inputenc}

\usepackage{lmodern}
\usepackage{xspace}                                     
\usepackage[normalem]{ulem}

\usepackage{amsfonts,amsmath,amssymb, amsthm, mathtools}
\usepackage{dsfont} 
\usepackage{algorithmicx,algpseudocode,algorithm}

\usepackage[usenames,dvipsnames,table]{xcolor}

\usepackage{csquotes}

\usepackage{relsize}

\usepackage{multirow}
\usepackage{chngpage} 
\usepackage{mfirstuc}

\usepackage{tikz}
\usetikzlibrary{arrows}
\usetikzlibrary{calc,decorations.pathmorphing,patterns}

\ifnum\withindex=1
  \usepackage{makeidx}
  \usepackage{ifthen}
  \newcommand\indexed[2][]{\ifthenelse{\equal{#1}{}}{#2\index{#2}}{#2\index{#1}}}
  \makeindex \fi

\usepackage[backref,colorlinks,citecolor=blue,bookmarks=true]{hyperref}
\usepackage{aliascnt}
\usepackage[numbered]{bookmark}

\usepackage{fullpage}

\usepackage[shortlabels]{enumitem}
  \setitemize{noitemsep,topsep=3pt,parsep=2pt,partopsep=2pt}   \setenumerate{itemsep=1pt,topsep=2pt,parsep=2pt,partopsep=2pt}
  \setdescription{itemsep=1pt}
  
\ifnum\withnotes=1
  \usepackage[colorinlistoftodos,textsize=scriptsize]{todonotes}
\fi

\usepackage{verbatim}

\usepackage{mleftright} 
\usepackage{titling}
 
\makeatletter
\newtheorem*{rep@theorem}{\rep@title}
\newcommand{\newreptheorem}[2]{\newenvironment{rep#1}[1]{ \def\rep@title{#2 \ref{##1}} \begin{rep@theorem}} {\end{rep@theorem}}}

\@ifundefined{theorem}{    \theoremstyle{plain}   	\newtheorem{theorem}{Theorem}[section]
  	\newreptheorem{theorem}{Theorem}
  	\newaliascnt{coro}{theorem}
  	  \newtheorem{corollary}[coro]{Corollary}
  	  \newreptheorem{corollary}{Corollary}
  	\aliascntresetthe{coro}
  	\newaliascnt{lem}{theorem}
  		\newtheorem{lemma}[lem]{Lemma}
  	\aliascntresetthe{lem}
  	\newaliascnt{clm}{theorem}
  		\newtheorem{claim}[clm]{Claim}
	\aliascntresetthe{clm}
	\newaliascnt{fct}{theorem}
 	 	\newtheorem{fact}[fct]{Fact}
	\aliascntresetthe{fct}
  	
  \newaliascnt{prop}{theorem}
  		
	\aliascntresetthe{prop}
	\newaliascnt{conj}{theorem}
  		
	\aliascntresetthe{conj}
	\newaliascnt{observ}{theorem}
  		\newtheorem{observation}[observ]{Observation}
	\aliascntresetthe{observ}
  \theoremstyle{remark}   \newaliascnt{rk}{theorem}
  		\newtheorem{remark}[rk]{Remark}
	\aliascntresetthe{rk}

  \theoremstyle{definition}   	\newaliascnt{defn}{theorem}
 		 \newtheorem{definition}[defn]{Definition}
 	 \aliascntresetthe{defn}
}{}
\makeatother

\providecommand{\email}[1]{\href{mailto:#1}{\nolinkurl{#1}\xspace}}

\ifnum\withcolors=1
  \newcommand{\new}[1]{{\color{red} {#1}}}               \newcommand{\inote}[1]{{\color{blue} {#1}}}   \newcommand{\lnote}[1]{{\color{Bittersweet} {#1}}}    \newcommand{\enote}[1]{{\color{ForestGreen} {#1}}}   \newcommand{\rnote}[1]{{\color{orange} {#1}}}   \newcommand{\cnote}[1]{{\color{RubineRed} {#1}}}

\else
  \newcommand{\new}[1]{{{#1}}}

  \newcommand{\inote}[1]{{ {#1}}}
  \newcommand{\lnote}[1]{{ {#1}}}
  \newcommand{\enote}[1]{{ {#1}}}
  \newcommand{\rnote}[1]{{ {#1}}}
  \newcommand{\cnote}[1]{{ {#1}}}

\fi

\ifnum\withnotes=1

\else

\fi
\newcommand{\ignore}[1]{\leavevmode\unskip}  
\newcommand{\eps}{\ensuremath{\varepsilon}\xspace}
     \newcommand{\eqdef}{\stackrel{\rm def}{=}}

\newcommand{\littleO}[1]{{o\mleft( #1 \mright)}}
\newcommand{\bigO}[1]{{O\mleft( #1 \mright)}}

\newcommand{\bigOmega}[1]{{\Omega\mleft( #1 \mright)}}
\newcommand{\tildeO}[1]{\tilde{O}\mleft( #1 \mright)}

\providecommand{\poly}{\operatorname*{poly}}

\newcommand{\setOfSuchThat}[2]{ \left\{\; #1 \;\colon\; #2\; \right\} }

\newcommand\restr[2]{{  \left.\kern-\nulldelimiterspace   #1   \vphantom{\big|}   \right|_{#2}   }}

\newcommand{\proba}{\Pr}
\newcommand{\probaOf}[1]{\proba\!\left[\, #1\, \right]}

\newcommand{\abs}[1]{\left\lvert #1 \right\rvert}

\newcommand{\flr}[1]{\left\lfloor #1 \right\rfloor}

\newcommand{\pdfsamp}{dual\xspace}
\newcommand{\cdfsamp}{cumulative dual\xspace}
\newcommand{\Pdfsamp}{\expandafter\capitalisewords\expandafter{\pdfsamp}}
\newcommand{\Cdfsamp}{\expandafter\capitalisewords\expandafter{\cdfsamp}}

\newcommand{\AND}{{\sf{}AND}\xspace}
\newcommand{\OR}{{\sf{}OR}\xspace}
\newcommand{\NOT}{{\sf{}NOT}\xspace}
\newcommand{\hypercube}{\{0,1\}^n}
\newcommand{\Boolean}{\hypercube \to \{0,1\}}
\DeclareMathOperator{\bias}{bias}
\DeclareMathOperator{\nstab}{Stab}

\newcommand{\expbias}[2]{\operatorname{ExpBias}_{#1}(#2)}
\newcommand{\parity}[1]{\PAR_{#1}}

\makeatletter

\newcommand{\Rom}[1]{\expandafter\@slowromancap\romannumeral #1@}

\makeatother

\makeatletter
\newcommand\ackname{Acknowledgements}
\if@titlepage
  \newenvironment{acknowledgements}{      \titlepage
      \null\vfil
      \@beginparpenalty\@lowpenalty
      \begin{center}        \bfseries \ackname
        \@endparpenalty\@M
      \end{center}}     {\par\vfil\null\endtitlepage}
\else
  
\fi
\makeatother
 
\usepackage{liyang}

\begin{document}

\title{Learning circuits with few negations}
\author{
  \enote{Eric Blais}\thanks{\email{eric.blais@uwaterloo.ca}, University of Waterloo. This work was completed as a Simons Postdoctoral Fellow at the Massachusetts Institute of Technology.}
  \and \cnote{Cl\'{e}ment L. Canonne}\thanks{\email{ccanonne@cs.columbia.edu}, Columbia University.}
  \and \inote{Igor C. Oliveira}\thanks{\email{oliveira@cs.columbia.edu}, Columbia University.}
  \and \rnote{Rocco A. Servedio}\thanks{\email{rocco@cs.columbia.edu}, Columbia University. Supported by NSF grants CCF-1115703 and CCF-1319788.}
  \and \lnote{Li-Yang Tan}\thanks{\email{liyang@cs.columbia.edu}, Columbia University.}
}
\makeatletter
  \hypersetup{pdftitle=\@title, pdfauthor={Eric Blais, Cl\'{e}ment Canonne, Igor C. Oliveira, Rocco A. Servedio and Li-Yang Tan}}
\makeatother

  \maketitle

\makeatletter{}
\begin{abstract}

Monotone Boolean functions, and the monotone Boolean circuits that compute them, have been intensively studied in complexity theory. In this paper we study the structure of Boolean functions in terms of the minimum number of negations in any circuit computing them, a complexity measure that interpolates between monotone functions and the class of all functions.  We study this generalization of monotonicity from the vantage point of learning theory, giving near-matching upper and lower bounds on the uniform-distribution learnability of circuits in terms of the number of negations they contain\ignore{(and completely independent of their size)}.  Our upper bounds are based on a new structural characterization of negation-limited circuits that extends a classical result of A. A. Markov. Our lower bounds, which employ Fourier-analytic tools from hardness amplification,
give new results even for circuits with no negations (i.e. monotone functions).

\end{abstract}
 
\clearpage
\makeatletter{}
\section{Introduction} \label{sec:intro}

A \emph{monotone} Boolean function $f\colon \{0,1\}^n \to \{0,1\}$ is one that satisfies
$f(x) \leq f(y)$ whenever $x \preceq y$, where $\preceq$ denotes the bitwise partial order on
$\{0,1\}^n$.
The structural and combinatorial properties of monotone Boolean functions have been intensively studied
for many decades, see e.g. \cite{Korshunov:03} for an in-depth survey.  Many famous results in circuit complexity deal with monotone functions, including celebrated lower bounds on monotone circuit size and monotone formula size (see e.g. \cite{RazWigderson:90,Razborov:85a} and
numerous subsequent works).

Monotone functions are also of considerable interest in computational learning theory, in particular with respect to the model of learning under the uniform distribution.  In an influential paper,
Bshouty and Tamon \cite{BshoutyTamon:96} showed that any monotone Boolean function $f\colon \{0,1\}^n \to
\{0,1\}$ can be learned from uniform random examples
to error $\eps$ in time $n^{O(\sqrt{n}/\eps)}$.  They also gave a lower bound,
showing that no algorithm running in
time $2^{cn}$ for any $c<1$ can learn arbitrary monotone functions to accuracy $\eps=1/
(\sqrt{n} \log n).$  (Many other works  in learning theory such as
\cite{Angluin:88,KearnsValiant:94,BBL:98,AmanoMaruoka:02,Servedio:04iandc,OdonnellServedio:07,OWimmer:09}
deal with learning monotone functions from a range of different perspectives
and learning models, but we limit our focus in this paper to learning to
high accuracy with respect to the uniform distribution.)

\subsection{Beyond monotonicity:  Inversion complexity, alternations, and Markov's theorem.}
Given the importance of monotone functions in complexity theory and
learning theory, it is natural to consider various generalizations of
monotonicity.
One such generalization arises from the simple observation that
monotone Boolean functions are precisely the functions computed
by \emph{monotone Boolean circuits}, i.e.  circuits which have only \AND
and \OR gates but no negations.
Given this, an obvious generalization of monotonicity is obtained by considering
functions computed by Boolean circuits
that have a small number of negation gates.  The
\emph{inversion complexity}
of $f\colon \{0,1\}^n \to \{0,1\}$, denoted $I(f)$,
is defined to be the minimum number of negation gates in any \AND/\OR/\NOT
circuit (with access to constant inputs 0/1) that computes $f$.
We write $\mathcal{C}_t^n$ to denote the class of $n$-variable
Boolean functions $f\colon \{0,1\}^n \to \{0,1\}$ that have
$I(f) \leq t$.

Another generalization of monotonicity is obtained by
starting from an alternate characterization of monotone Boolean functions.
A function $f\colon \{0,1\}^n \to \{0,1\}$ is monotone if and only if the
value of $f$ ``flips'' from 0 to 1 at most once as the input
$x$ ascends any chain in $\{0,1\}^n$ from $0^n$ to $1^n$.
(Recall that a chain of length $\ell$ is an increasing sequence $(x^1,
\ldots, x^\ell)$ of vectors in $\hypercube$, i.e. for every $j \in
\{1,\dots,\ell-1\}$ we have $x^j \prec x^{j+1}$.)
Thus, it is natural to consider a generalization of monotonicity that allows
more than one such ``flip'' to occur.  We make this precise with the following
notation and terminology:  given a Boolean function $f\colon \Boolean$
and a chain $X = (x^1, \ldots, x^\ell)$, a position $j \in [\ell -1]$ is
said to be \emph{alternating}
with respect to $f$ if $f(x^j) \neq f(x^{j+1}).$
We write $A(f,X) \subseteq [\ell -1]$ to denote the set of
alternating positions in $X$ with respect to $f$, and we let $a(f,X)
=\abs{ A(f,X) }$ denote its size.
We write $a(f)$ to denote the maximum of $a(f,X)$ taken over all
chains $X$ in $\{0,1\}^n$, and we say that $f\colon \{0,1\}^n \to \{0,1\}$ is
\emph{$k$-alternating} if $a(f) \leq k$.

A celebrated result of A. A. Markov from 1957 \cite{Markov:57} gives a tight
quantitative connection between the inversion and alternation complexities defined above:
\newtheorem*{Markov}{Markov's Theorem} \label{thm:markov}
\begin{Markov}
Let $f\colon \{0,1\}^n \to \{0,1\}$ be a function which is not identically 0.  Then
(i) if $f(0^n)=0$, then $I(f) =  \lceil \log (a(f)+1) \rceil -1$; and
(ii) if $f(0^n)=1$, then $I(f) = \lceil \log (a(f)+2) \rceil - 1.$
\end{Markov}

This robustness
motivates the study of circuits which contain few negation gates,
and indeed such circuits have been studied in complexity theory.
Amano and Maruoka \cite{AmanoMaruoka:clique} have given bounds on the
computational power of such circuits, showing that circuits for the
clique function which contain fewer than ${\frac 1 6} \log \log n$
many negation gates must have superpolynomial size.
Other works have studied the effect of limiting the number of
negation gates in formulas \cite{Morizumi09}, bounded-depth
circuits \cite{ST02}, constant-depth circuits \cite{SW93}
and non-deterministic circuits \cite{Morizumi09nondet}.
In the present work, we study circuits with few negations
from the vantage point of computational
learning theory, giving both positive and negative results.

\subsection{Our results}
We begin by studying the structural properties of functions that are computed
or approximated by circuits with few negation gates. In \autoref{sec:structural}
 we establish the following extension of Markov's theorem:

\begin{theorem} \label{thm:structural}
Let $f$ be a $k$-alternating Boolean function. Then $f(x) = h(m_1(x),\ldots,m_k(x))$, where each $m_i(x)$ is monotone and $h$ is either the parity function or its negation.  {Conversely, any function of this form is $k$-alternating.}
\end{theorem}

\autoref{thm:structural} along with Markov's theorem yields the following characterization of $\mathcal{C}^n_t$: 

\begin{corollary}
\label{cor:circuit}
Every $f \in {\cal C}^n_t$ can be expressed as $f = h(m_1,\dots,m_T)$ where $h$ is either
$\PAR_T$ or its negation, each $m_i \colon \{0,1\}^n \to \{0,1\}$
is monotone, and $T = O(2^t).$
\end{corollary} 

A well-known consequence of Markov's theorem is that every Boolean
function is exactly computed by a circuit which has only $\log n$
negation gates, and 
{as we shall see} an easy argument shows that every Boolean function
is {$0.01$-approximated} by a circuit with {$\frac{1}{2}\log n + \bigO{1}$}
negations. In \autoref{sec:structural} we note that
no significant savings are possible over this easy upper bound: \begin{theorem}\label{prop:approx:random:lb:negations}
For almost every function $f\colon \{0,1\}^n \to \{0,1\}$,
any Boolean circuit $C$ that $0.01$-approximates $f$ must contain {${\frac 1 2}\log n-\bigO{1}$} negations.
\end{theorem}

We then turn to our main topic of investigation, the uniform-distribution
learnability of circuits with few negations.
We use our new extension of Markov's theorem, \autoref{thm:structural}, 
{to obtain a generalization of the Fourier-based uniform-distribution learning algorithm of Bshouty and Tamon \cite{BshoutyTamon:96} for monotone circuits:}

\begin{theorem} \label{thm:learn-upper}
There is a uniform-distribution learning algorithm which
learns any unknown \mbox{$f\in{\cal C}^n_t$} {from random examples}
to error $\eps$ in time $n^{O(2^t \sqrt{n}/\eps)}.$
\end{theorem}

\autoref{thm:learn-upper} immediately leads to
the following question:
can an even faster learning algorithm be given for
circuits with $t$ negations, or is the running time of
\autoref{thm:learn-upper} essentially the best possible?
Interestingly, prior to our work a matching lower bound for
\autoref{thm:learn-upper} was not known even for the
special case of monotone functions (corresponding to $t=0$).
As mentioned earlier, Bshouty and Tamon proved that to achieve
accuracy $\eps = 1 / (\sqrt{n} \log n)$ any learning
algorithm needs time $\omega(2^{cn})$ for any $c < 1$ (see \autoref{fact:bt:hardness} for a slight sharpening of this statement).
For larger values of $\eps$, though, the strongest previous
lower bound was due to Blum, Burch and Langford
\cite{BBL:98}.  Their Theorem~10
implies that any membership-query algorithm
that learns monotone functions to error $\eps < {\frac 1 2} - c$
(for any $c > 0$) must run in time
$2^{\Omega(\sqrt{n})}$ (in fact, must make at least this many
membership queries).  However, this lower bound does not
differentiate between the number of membership queries
required to learn to high accuracy versus ``moderate'' accuracy -- say,
{$\eps=1/n^{1/10}$ versus $\eps=1/10$}.  Thus the following question
was unanswered prior to the current paper:  what is the best lower bound
that can be given, both as a function of $n$ and $\eps$,
on the complexity of learning monotone functions to accuracy $\eps$?

We give a fairly complete answer to this question, providing a lower
bound as a function of $n,\eps$ and $t$ on the complexity
of learning circuits with $t$ negations.  Our lower bound essentially
matches the upper bound of \autoref{thm:learn-upper}, and
is thus simultaneously essentially optimal in all three parameters
$n,\eps$ and $t$ for a wide range of settings of $\eps$ and $t$.
Our lower bound result is the following:

\begin{theorem} \label{thm:lower}
For any $t \leq {\frac 1 {28}} \log n$ and any $\eps \in [1/n^{1/12},1/2 - c]$, $c>0$, any membership-query algorithm that learns any
unknown function $f \in {\cal C}^n_t$ to error $\eps$
must make $2^{\Omega(2^t \sqrt{n}/\eps)}$ membership queries.
\end{theorem}

We note that while our algorithm uses only uniform random examples, our lower
bound holds even for the stronger model in which the learning algorithm is allowed
to make arbitrary membership queries on points of its choosing.

\autoref{thm:lower} is proved using tools from the study of
hardness amplification. The proof involves a few steps. We start with a strong lower bound for the task of learning to high accuracy the class of balanced monotone Boolean functions (reminiscent of the lower bound obtained by Bshouty and Tamon). Then we combine hardness amplification techniques and results on the noise sensitivity of monotone functions in order to get stronger and more general lower bounds for learning monotone Boolean functions to moderate
accuracy. Finally, we use hardness amplification once more to lift this result into a lower bound for learning circuits with few negations to moderate accuracy. 
An ingredient employed in this last stage is to use a $k$-alternating combining function which ``behaves like'' the parity function
on (roughly) $k^2$ variables; this is crucial in order for us to obtain our essentially optimal final lower bound of
$2^{\Omega(2^t \sqrt{n}/\eps)}$ for circuits with $t$ negations. These results are discussed in more detail in \autoref{section:learning:strong:lb}.
  
\makeatletter{}
\section{Structural facts about computing and approximating
functions with low inversion complexity}
\label{sec:structural}

\subsection{An extension of Markov's theorem.} 
We begin with the proof of our new extension of Markov's theorem.  For any $A \subseteq \{0,1\}^n$ let $\ind[A]: \{0,1\}^n \to \{0,1\}$ be the characteristic function of $A$.  For $f\colon \{0,1\}^n \to \{0,1\}$ and $x \in \{0,1\}^n$, we write $a_f(x)$ to denote
\[ a_f(x) \eqdef \max\{a(f,X): X \text{ is a chain that starts at $x$}\},\]
and note that $a(f) = \max_{x\in\zo^n} \{ a_f(x)\} = a_f(0^n)$.  
For $0 \leq \ell \leq a(f)$ let us write $S^f_\ell$ to denote
$ S^f_\ell \eqdef\{x \in \{0,1\}^n: a_f(x)=\ell\}$, and let $T^f_\ell \eqdef S^f_0 \cup \cdots \cup 
S^f_\ell.$  We note that $S^f_1,\ldots,S^f_{a(f)}$ partition the set of all inputs: $S_i^f \cap S_j^f = \emptyset$ for all $i \neq j$, and $T_{a(f)}^f = S_1^f \cup\cdots\cup S_{a(f)}^f = \zo^n$. \medskip 

We will need the following simple observation:

\begin{observation} 
\label{obs:useful}
Fix any $f$ and any $x \in \{0,1\}^n$.  
If $x \in S^f_\ell$ and $y \succ x$ then $y \in S^f_{\ell^\prime}$ for some $\ell^\prime \leq \ell.$ {Furthermore, if $f(y) \ne f(x)$ then $\ell^\prime < \ell$. }
\end{observation}

\begin{reptheorem}{thm:structural} (Restated)
Fix $f\colon \{0,1\}^n \to \{0,1\}$ and let $k \eqdef a(f).$  Then $f = h\big(\ind\big[T^f_0\big],\dots,\ind\big[T^f_{k-1}\big]\big)$, where 
\begin{itemize}
\item[(i)] the functions $\ind\big[T^f_\ell\big]$ are monotone for all $0\le \ell\le k$, 
\item[(ii)] $h : \zo^k\to\zo$ is $\PAR_k$ if $f(0^n)=0$ and $\neg\, \PAR_k$ if $f(0^n) = 1$, 
\end{itemize} 
and $\PAR_k(x) = x_1 \oplus \cdots \oplus x_k$ is the parity function on $k$ variables.
Conversely, for any monotone Boolean functions $m_1,\dots,m_k$, any Boolean function of the form $h(m_1,\dots,m_k)$    
is $k$-alternating.
\end{reptheorem} 

\begin{proof}
Claim (i) follows immediately from \autoref{obs:useful} above.
The proof of (ii) is by induction on $k$.  In the base case $k=0$, we have that $f$
is a constant function and the claim is immediate.

For the inductive step, suppose that the claim holds for all functions $f^\prime$ that have $a({f^\prime}) \leq k-1$.
We define $f^\prime\colon \{0,1\}^n \to \{0,1\}$ as $f^\prime = f \oplus \ind\big[S^f_k\big].$
\autoref{obs:useful} implies that $S^{f^\prime}_\ell = S^f_\ell$ for all $0\le \ell \le k-2$ and $S^{f^\prime}_{k-1} = S^f_{k-1}\cup S^f_k$, and in particular, $a(f) = k-1$.  Therefore we may apply the inductive hypothesis to $f^\prime$ and express it
as $f^\prime = h'\big(\ind\big[T^{f^\prime}_0\big],\dots,\ind\big[T^{f^\prime}_{k-2}\big]\big).$  Since $T^{f^\prime}_\ell=T^{f}_\ell$ for $0 \leq \ell \leq k-2$, we may use this along with the
fact that $\ind\big[S^f_k\big]=\neg\,\ind\big[T^{f}_{k-1}\big]$ to get: 
\[
f=f^\prime \oplus \ind\big[S^f_k\big] = 
h'\big(\ind\big[T^{f^\prime}_0\big],\dots,\ind\big[T^{f^\prime}_{k-2}\big]\big)  \oplus \neg\,\ind\big[T^{f}_{k-1}\big] = 
h'\big(\ind\big[T^{f}_0\big],\dots,\ind\big[T^{f}_{k-2}\big]\big)  \oplus \neg\,\ind\big[T^{f}_{k-1}\big]
\]
and the inductive hypothesis holds (note that $0^n \in S_k^{f}$).

The converse is easily verified by observing that any chain in $\{0,1\}^n$ can induce at most $k+1$ possible vectors of values for
$(m_1,\dots,m_k)$ because of their monotonicity.
\end{proof}

\autoref{thm:structural} along with Markov's theorem immediately yields the following corollary: 

\begin{repcorollary}{cor:circuit}
Every $f \in {\cal C}^n_t$ can be expressed as $f = h(m_1,\dots,m_T)$ where $h$ is either
$\PAR_T$ or its negation, each $m_i \colon \{0,1\}^n \to \{0,1\}$
is monotone, and $T = O(2^t).$
\end{repcorollary}

\subsection{Approximation.}

As noted earlier, Markov's theorem implies that every $n$-variable Boolean function can be
exactly computed by a circuit with (essentially) $\log n$ negations (since $a(f) \leq n$
for all $f$).  If we set a less ambitious goal of \emph{approximating} Boolean functions
(say, having a circuit correctly compute $f$ on a $1-\eps$ fraction of all $2^n$ inputs),
can significantly fewer negations suffice?

We first observe that every Boolean function $f$ is 
$\eps$-close (with respect to the uniform distribution) to a function $f^\prime$
that has $a(f^\prime) \leq O(\sqrt{n \log 1/\eps})$.  The function $f^\prime$
is obtained from $f$ simply by setting $f^\prime(x)=0$ for all inputs $x$ that have
Hamming weight outside of $[n/2 - O(\sqrt{n \log 1/\eps}),
n/2 + O(\sqrt{n \log 1/\eps})]$; a standard Chernoff bound implies that $f$
and $f^\prime$ disagree on at most $\eps 2^n$ inputs.  Markov's theorem then implies that
the inversion complexity $I(f^\prime)$ is at most $\frac{1}{2}(\log n + \log \log {\frac 1 \eps}) + O(1)$.
Thus, every Boolean function can be approximated to high accuracy by a circuit with only $\frac{1}{2}
\log n + O(1)$ negations.

We now show that this upper bound is essentially optimal:  for almost every
Boolean function, any $0.01$-approximating circuit must contain at least $\frac{1}{2} \log n
- O(1)$ negations.  To prove this, we recall the notion of the 
\emph{total influence} of a Boolean function $f$:  this is
\[
\Inf[f] = \sum_{i=1}^n \Inf_i[f],
\quad \text{where} \quad
\Inf_i[f] = \Pr_{x \in \{0,1\}^n}[f(x) \neq f(x^{\oplus i})]
\]
and $x^{\oplus i}$ denotes $x$ with its $i$-th coordinate flipped.  The total influence of $f$ is easily seen to equal $\alpha n$, where $\alpha \in [0,1]$ is the fraction of
all edges $e=(x,x')$ in the Boolean hypercube that are bichromatic, i.e. have $f(x) \neq
f(x').$  In Appendix~\ref{ap:proof-influence} we prove the following lemma:

\new{
\begin{lemma}\label{claim:high:influence:alternations}
Suppose $f\colon\Boolean$ is such that ${\Inf[f]}=\bigOmega{n}$. Then $a(f) = \bigOmega{\sqrt{n}}$.
\end{lemma}
}

It is easy to show that a random function has influence $\frac{n}{2}(1-\littleO{1})$ with probability $1-2^{-n}$. Given this, \autoref{claim:high:influence:alternations}, together with the elementary fact that if $f^\prime$ is $\eps$-close to $f$ then $|\Inf(f^\prime) - \Inf(f)| \leq 2 \eps n$, directly yields the following:
\begin{reptheorem}{prop:approx:random:lb:negations}With probability $1-2^{-n}$, any $0.01$-approximator $f^\prime$ for a random function $f$ must have 
inversion complexity $I(f^\prime) \geq \frac{1}{2} \log n - O(1).$
\end{reptheorem} 

\begin{remark}\label{rk:happy:with:result} {The results in this section (together with simple information-theoretic arguments showing that random functions are hard to learn) imply that one cannot expect to have a learning algorithm \new{(even to constant accuracy)} for the class} ${\cal C}^{n}_{\frac{1}{2}\log n + O(1)}$
of circuits with $\frac{1}{2} \log n + O(1)$ negations in time significantly better than $2^n$.  As we shall see in \autoref{section:learning:strong:ub}, for any fixed $\delta > 0$ it
\emph{is} possible to learn ${\cal C}^n_{(\frac{1}{2} - \delta) \log n}$ to accuracy $1-\eps$  in time $2^{\tildeO{n^{1-\delta}}/\eps}$.
\end{remark}
 
\makeatletter{}
\section{Learning circuits with few negations}\label{s:learnability}

\subsection{A learning algorithm for \texorpdfstring{${\mathcal C}^n_t$}{C(n,t)}.}
\label{section:learning:strong:ub}

We sketch the learning algorithm and analysis of Bshouty and Tamon 
\cite{BshoutyTamon:96}; using the results from \autoref{sec:structural}
our \autoref{thm:learn-upper} will follow easily from their approach.
Our starting point is the simple observation that functions with
good ``Fourier concentration'' can be learned to high accuracy under the
uniform distribution simply by estimating
all of the low-degree Fourier coefficients.  This fact, established
by Linial, Mansour and Nisan, is often referred to as
the ``Low-Degree Algorithm:''

\begin{theorem}[Low-Degree Algorithm (\cite{LMN:93})] \label{thm:lda}
Let $\mathcal{C}$ be a class of Boolean functions such that 
for $\eps > 0$ and $\tau=\tau(\eps,n)$, 
\[
\sum_{|S| > \tau} \hat{f}(S)^2 \leq \eps
\]
for any $f\in\mathcal{C}$. Then $\mathcal{C}$ can be learned from
uniform random examples in time $\poly(n^\tau,1/\eps)$.
\end{theorem}

Using the fact that every monotone function $f\colon \{0,1\}^n \to \{0,1\}$
has total influence $\Inf(f) \leq \sqrt{n}$, and the well-known
Fourier expression $\Inf(f) = \sum_{S} \hat{f}(S) \cdot |S|^2$ for
total influence, a simple application of Markov's inequality let
Bshouty and Tamon show that every monotone function $f$
has
\[
\sum_{|S| > \sqrt{n}/\eps} \hat{f}(S)^2 \leq \eps.
\]
Together with \autoref{thm:lda}, this gives their
learning result for monotone functions.

Armed with \autoref{cor:circuit}, it is straightforward to extend
this to the class ${\mathcal C}^n_t$. \autoref{cor:circuit}
and a union bound immediately give that every $f \in {\mathcal C}^n_t$
has $\Inf(f) \leq O(2^t) \sqrt{n}$, so the Fourier 
expression for influence  and Markov's inequality give that
\[
\sum_{|S| > O(2^t) \sqrt{n}/\eps} \hat{f}(S)^2 \leq \eps
\]
for $f \in {\mathcal C}^n_t.$
\autoref{thm:learn-upper} follows immediately using the Low-Degree
Algorithm.\medskip

\ignore{
\begin{remark}
{\huge FINISH}
We note that this bound is nontrivial only for $t < \frac{1}{2}\log n$;
this was to be expected, as having even $\frac{1}{2}\log n$ negation gates allows us to approximate \emph{any} Boolean function.
\end{remark}  
}  

An immediate question is whether this upper bound on the complexity
of learning $\mathcal{C}^n_t$ is  {optimal}; we give an affirmative answer
in the next subsection.

\subsection{Lower bounds for learning.}\label{section:learning:strong:lb}

As noted in the introduction, we prove information-theoretic lower
bounds against learning algorithms that make a limited number of
membership queries.
We start by establishing a new lower bound on the number of membership
queries that are required to learn \emph{monotone} functions to
high accuracy, and then build on this to provide a lower bound 
for learning ${\mathcal C}^n_t.$  Our query lower bounds are essentially tight,
matching the upper bounds (which hold for learning from uniform random
examples) up to logarithmic factors in the exponent.

We first state the results; the proofs are deferred to \autoref{section:learning:strong:lb:proofs}.
We say that a Boolean function $f$ is \emph{balanced} if 
$\Pr_x[f(x)=0]=\Pr_x[f(x)=1]=1/2.$

\begin{theorem}\label{theorem:strong:learn:lb:monotone}
There {exists} a class $\mathcal{H}$ of balanced $n$-variable
monotone Boolean functions such that
for any $\eps\in[\frac{1}{n^{1/6}}, 1/2-c]$, $c>0$, 
learning $\mathcal{H}_n$ to accuracy $1-\eps$ requires 
$2^{\bigOmega{\sqrt{n}/\eps}}$ membership queries.
\end{theorem}

This immediately implies the following corollary, which {essentially} closes the 
gap in our understanding of the hardness of learning monotone functions:
\begin{corollary}\label{corollary:strong:learn:lb:monotone}
For any $\eps=\bigOmega{1/{n^{1/6}}}$ \new{bounded away from $1/2$},
learning $n$-variable monotone functions to accuracy $1-\eps$ 
requires $2^{\tilde{\Theta}(\sqrt{n})/\eps}$ queries.
\end{corollary}

Using this class $\mathcal{H}$ as a building block, we obtain 
the following hardness of learning result for the class of $k$-alternating
functions:
\begin{theorem}\label{theorem:strong:learn:lb:kmonotone}
For any function $k\colon\N\to\N$, there exists a class $\mathcal{H}^{(k)}$ 
of balanced $k=k(n)$-alternating 
$n$-variable Boolean functions such that, for any $n$ sufficiently large 
and $\eps > 0$ such that (i) $2\leq k < n^{1/14}$, and (ii) 
$k^{7/3}/n^{1/6} \leq \eps \leq \frac{1}{2} - c$,
learning $\mathcal{H}^{(k)}$ to accuracy $1-\eps$ requires $2^{\bigOmega{k\sqrt{n}/\eps}}$ membership queries.
\end{theorem}

\noindent
(We note that the tradeoff between the ranges of $k$ and $\eps$ that is
captured by condition (ii) above seems to be inherent to our approach
and not a mere artifact of the analysis; 
see \autoref{remark:lb:against:kmonotone:tradeoff:k:eps}.)
This theorem immediately yields the following:

\begin{corollary}\label{corollary:strong:learn:lb:kmonotone}
Learning the class of $k$-alternating functions to accuracy $1-\eps$ 
in the uniform-distribution membership-query model 
requires $2^{\bigOmega{k\sqrt{n}/\eps}}$ membership queries, 
for any $k=\bigO{n^{1/28}}$ and  $\eps\in [1/{n^{1/12}},\frac{1}{2} - c]$.
\end{corollary}

\begin{corollary}\label{corollary:strong:learn:lb:circuits:few:neg}
For $t \leq \frac{1}{28} \log n$, learning $\mathcal{C}_t^n$ to 
accuracy $1-\eps$ requires $2^{\bigOmega{2^t\sqrt{n}/\eps}}$ membership 
queries, for any $\eps \in [2^{7t/3}/n^{1/6},\frac{1}{2} - c]$.
\end{corollary}

\subsubsection{Proofs.}\label{section:learning:strong:lb:proofs}

We require the following standard notion of \emph{composition} 
for two functions $f$ and $g$:

\begin{definition}[Composition]
For $f\colon\{0,1\}^m\to\{0,1\}$ and $g\colon\{0,1\}^r\to\{0,1\}$, 
we denote by $g\otimes f$ the Boolean function on $n=mr$ inputs defined by
\[
    (g\otimes f)(x) \eqdef g( \underbrace{f,\dots,f}_{r} )(x) = g( f(x_1,\dots,x_m), \dots, f(x_{(r-1)m+1},\dots,x_{rm}) )
\]
Similarly, for any $g\colon\{0,1\}^r\to\{0,1\}$ and $\mathcal{F}_m$ a 
class of Boolean functions on $m$ variables, we let
\[
g\otimes \mathcal{F}_m = \setOfSuchThat{g\otimes f}{f\in\mathcal{F}_m}
\]
and $g\otimes \mathcal{F} = \{ g\otimes \mathcal{F}_m \}_{m\geq 1}$.
\end{definition}

\medskip \noindent {\bf Overview of the arguments.}
Our approach is based on hardness amplification. 
In order to get our lower bound against  learning $k$-alternating 
functions, we \textsf{(a)} start from a lower bound ruling out very high-accuracy learning of \emph{monotone} functions;
 \textsf{(b)} use a suitable monotone combining function to get an \textsf{XOR}-like hardness amplification, 
yielding a lower bound for learning (a subclass of) monotone functions to moderate accuracy; \textsf{(c)} repeat this approach on this subclass with a 
different (now $k$-alternating) combining function to obtain our final lower 
bound, for learning $k$-alternating functions to moderate accuracy.\begin{equation} \label{eq:diagram}
    \underbracket[0.5pt]{\begin{bmatrix}
      \text{{high-accuracy}}\\
      \text{monotone}
    \end{bmatrix}}_{\sf(a)} 
    \xrightarrow[\vphantom{k\text{-}}\text{monotone}]{\quad\bigotimes
\text{-like}\quad}
    \underbracket[0.5pt]{\begin{bmatrix}
      \text{moderate accuracy}\\
      \text{monotone}
    \end{bmatrix} }_{\sf(b)} 
    \xrightarrow[k\text{-alternating}]{\quad\bigotimes\text{-like}\quad}
    \underbracket[0.5pt]{\begin{bmatrix}
      \text{moderate accuracy}\\
      k\text{-alternating}
    \end{bmatrix} }_{\sf(c)} 
\end{equation}

In more detail, in both steps \textsf{(b)} and \textsf{(c)} 
the idea is to take as base functions the hard class from the previous step 
(respectively ``monotone hard to {learn to high accuracy}'', and ``monotone hard to 
learn to moderate accuracy''), and compose them with a very noise-sensitive function in 
order to amplify hardness.
Care must be taken to ensure that the combining function satisfies several 
necessary constraints (being monotone for \textsf{(b)} and 
$k$-alternating for \textsf{(c)}, and being as sensitive 
as possible to the correct regime of noise in each case).

\paragraph*{Useful tools.}
We begin by recalling a few notions and results that play a crucial role 
in our approach. 
\ignore{Note that for consistency, we phrase all theorems and definitions 
used in terms of 0/1-valued Boolean functions (instead of the somewhat more 
common equivalent $\pm 1$ notation).}

\begin{definition}[Noise stability]
For $f\colon\{0,1\}^n\to\{0,1\}$, the \emph{noise stability} of $f$ at $\eta\in[-1,1]$ is
\[
\nstab_\eta(f) \eqdef 1-2\probaOf{f(x)\neq f(y)} \]
where $x$ is drawn uniformly at random from $\{0,1\}^n$ and 
{$y$ is obtained from $x$ by independently for each bit having
$\Pr[y_i = x_i] = (1 + \eta)/2$ 
(i.e., $x$ and $y$ are \emph{$\eta$-correlated}).}
\end{definition}

\begin{definition}[Bias and expected bias]
The \emph{bias} of a Boolean function $h\colon\{0,1\}^n\to \{0,1\}$ is the 
quantity $\bias(h) \eqdef \max(\probaOf{h=1}, \probaOf{h=0})$, while 
the \emph{expected bias of $h$ at $\delta$} is defined as 
\mbox{$\expbias{\delta}{h}\eqdef {\E_{\rho}[\bias(h_\rho})]$}, where $\rho$ is 
a random restriction on $k$ coordinates where each coordinate is 
independently  left free with probability $\delta$ and 
set to 0 or 1 with same probability $(1-\delta)/2$.
\end{definition}

\begin{fact}[Proposition 4.0.11 from~\cite{ODThesis}]\label{fact:relation:noisestab:expbias}
For $\delta\in[0,1/2]$ and $f\colon\{0,1\}^n\to\{0,1\}$, we have
\[
    \frac{1}{2} + \frac{1}{2}\nstab_{1-2\delta}(f) \leq \expbias{2\delta}{f} \leq \frac{1}{2} + \frac{1}{2}\sqrt{\nstab_{1-2\delta}(f)}.
\]
\end{fact}

Building on Talagrand's probabilistic construction \cite{Talagrand:96} 
of a class of functions that are sensitive to very small noise, 
Mossel and O'Donnell~\cite{MO03} gave the following noise stability
upper bound.  (We state below a slightly generalized version of 
their Theorem~3, which follows from their proof with some minor changes; see Appendix~\ref{sec:MO-changes} for details of 
these changes.)

\begin{theorem}[Theorem 3 of~\cite{MO03}]\label{theo:odonnell:function}
There exists an absolute constant $K$ and an infinite 
family of \emph{balanced} monotone functions 
\mbox{$g_r\colon\{0,1\}^r\to\{0,1\}$} such that
$\nstab_{1-{\tau}/{\sqrt{r}}}(g_r)\leq 1-K\tau$ holds
for all sufficiently large $r$,
as long as $\tau \in \left[16/\sqrt{r},1\right]$.
\end{theorem}

Applying \autoref{fact:relation:noisestab:expbias}, it follows that
for the Mossell-O'Donnell function $g_r$ on $r$ inputs and any $\tau$ as above,
we have
\begin{equation}\label{eq:expbias:mod:function}
  \frac{1}{2} \leq \expbias{ \gamma }{g_r} \leq \frac{1}{2} + \frac{1}{2}\sqrt{1-K\tau} \leq 1-\frac{K}{4}\tau
\end{equation}
for $\gamma\eqdef \frac{\tau}{\sqrt{r}}$.\smallskip

We will use the above upper bound on expected bias together with the following
key tool from \cite{FLS11}, which gives a hardness amplification result for
uniform distribution learning.  This result builds on the original hardness
amplification ideas of O'Donnell \cite{ODThesis}.  
(We note that the original theorem statement from \cite{FLS11} deals
with the running time of learning algorithms, but inspection of the
proof shows that the theorem also applies to the number
of membership queries that the learning algorithms perform.)

\begin{theorem}[Theorem 12 of~\cite{FLS11}]\label{theo:rocco:xor:lemma}
Fix $g\colon\{0,1\}^r\to\{0,1\}$, and let $\mathcal{F}$ be a class of 
$m$-variable Boolean functions such that for every $f\in\mathcal{F}$, 
$\bias(f) \leq \frac{1}{2} + \frac{\epsilon}{8r}$. Let $A$ be a uniform 
distribution membership query algorithm that learns $g\otimes\mathcal{F}$ 
to accuracy $\expbias{\gamma}{g}+\epsilon$ using $T(m,r,1/\epsilon,1/\gamma)$ 
queries. Then there exists a uniform-distribution membership query 
algorithm $B$ that learns $\mathcal{F}$ to accuracy $1-\gamma$ using
$\bigO{T\cdot \poly({m}, r,1/\epsilon,1/\gamma)}$ membership queries.
\end{theorem}

\medskip
\noindent {\bf Hardness of learning monotone functions to high
accuracy.}
At the bottom level, corresponding to step \textsf{(a)} in \eqref{eq:diagram},
our approach relies on the following simple claim which 
states that monotone functions are hard to learn to very high accuracy.
(We view this claim, as essentially folklore; as noted in the introduction
it slightly sharpens a lower bound given in \cite{BshoutyTamon:96}.
A proof is given for completeness in Appendix~\ref{ap:proof-fact-bt-hardness}.)

\begin{claim}[A slice of hardness]\label{fact:bt:hardness}
There exists a class of balanced monotone Boolean functions 
$\mathcal{G}=\{\mathcal{G}_m\}_{m\in \N}$ and a universal constant $C$ 
such that, for any constants $0 < \alpha \leq 1/10$, 
learning $\mathcal{G}_m$ to error $0 < \eps \leq {\alpha}/{\sqrt{m}}$ 
requires at least $2^{Cm}$ membership queries.
\end{claim}

We now prove \autoref{theorem:strong:learn:lb:monotone}, i.e. we
establish a {stronger lower bound (in terms of the range of accuracy it applies to)} against learning the class of 
\emph{monotone} functions. We do this by amplifying the hardness 
result of \autoref{fact:bt:hardness} by composing the ``mildly hard'' class 
of functions $\mathcal{G}$ with a monotone function $g$ --- 
the Mossel-O'Donnell function of \autoref{theo:odonnell:function} 
--- 
that is 
very sensitive to small noise (intuitively, the noise rate here is comparable
to the error rate from \autoref{fact:bt:hardness}).
\ignore{Our hard class of functions will 
be compositions of the form ${g\otimes\mathcal{G}_m}$, where $g$ has roughly 
the same desirable properties as $\parity{}$ (very low noise stability, 
or equivalently expected bias very close to $1/2$). For this purpose, 
it is reasonable to consider for $g$ the Mossel-O'Donnell function 
of \autoref{theo:odonnell:function}.
}

\begin{proof}[Proof of \autoref{theorem:strong:learn:lb:monotone}]
We will show that there exists an absolute constant $\alpha > 0$ such that 
for any $n$ sufficiently large and $\tau \in [\frac{1}{n^{1/6}},1/2-c]$, 
there exist $m=m(n)$, $r=r(n)$ (both of which are $\omega_n(1)$) such that 
learning the class of (balanced) 
functions $\mathcal{H}_n={g_r\otimes\mathcal{G}_m}$ on $n=mr$ variables 
to accuracy $1-\tau$ requires at least $2^{\alpha\sqrt{n}/\tau}$ 
membership queries.

By contradiction, suppose we have an algorithm $A$ which, for all 
$m,r,\tau$ as above, learns the class $\mathcal{H}_n$ 
to accuracy $1-\tau$ using $T=T_A(n, \tau) < 2^{\alpha\sqrt{n}/\tau}$
membership queries.
We show that this implies that for infinitely many values of $m$, one can learn ${\mathcal{G}_m}$ to error $\eps=.1/\sqrt{m}$ with $2^{\littleO{m}}$ 
membership queries, in contradiction to \autoref{fact:bt:hardness}.

Fix any $n$ large enough and 
$\tau\in[\frac{1}{n^{1/6}}, .1]$,
and choose $m,r$ satisfying $mr=n$ and
$
\frac{5}{K}\cdot\frac{\tau}{\sqrt{r}} = \frac{.1}{\sqrt{m}},$
where $K$ is the constant from \autoref{theo:odonnell:function}. 
Note that this implies $m=\frac{K}{50}\cdot\frac{\sqrt{n}}{\tau} 
\in [\Theta(n^{1/2}),\Theta(n^{2/3})]$ so indeed both $m$ and $r$
are $\omega_n(1).$
(Intuitively, the value ${\frac {.1}{\sqrt{m}}}$ 
is the error we \emph{want} to achieve to get a contradiction, 
while the value 
$\frac{5}{K}\cdot\frac{\tau}{\sqrt{r}}$
is the error we \emph{can} get from \autoref{theo:rocco:xor:lemma}.) 
Note that we indeed can use the Mossel-O'Donnell function
from \autoref{theo:odonnell:function}, which requires $\tau > 
\frac{16}{\sqrt{r}}$ -- for our choice of $r$, this is equivalent to 
$\tau >  \Big(\frac{16\sqrt{K}}{\sqrt{50}}\Big)^{2/3}\frac{1}{n^{1/6}}$.
Finally, set $\eps\eqdef .1/\sqrt{m}$.
 
We apply \autoref{theo:rocco:xor:lemma} with $g\eqdef g_r$, 
$\gamma=(5/K)\tau/\sqrt{r}$ and $\epsilon=\tau/4$. 
(Note that all functions in $\mathcal{G}_m$ are balanced, 
and thus trivially satisfy the condition that 
$\bias(f)\leq \frac{\epsilon}{8r}$, and
recall that $1-\gamma$ is the accuracy the theorem guarantees against the 
original class $\mathcal{G}_m$.)  With these parameters we have
\ignore{
}
\begin{align*}
  \expbias{\gamma}{g} + \epsilon &\operatorname*{\leq}_{\text{Eq.}
\eqref{eq:expbias:mod:function}} 1-\frac{K}{4}\frac{5\tau}{K} + \frac{\tau}{4} 
= 1 - \tau \leq \operatorname{accuracy}(A).
\end{align*}
\autoref{theo:rocco:xor:lemma} gives that 
there exists a learning algorithm $B$ learning $\mathcal{G}_m$ to 
accuracy $1-\gamma \geq 1-\eps$ with $T_B=\bigO{ T\cdot 
\poly(m,r,1/\tau,1/\gamma) }=\bigO{ T\cdot \poly(n,1/\tau) }$ 
membership queries, that is,
$T_B = T_A(n,\tau)\cdot \poly(n,1/\tau) < 
2^{\alpha\sqrt{n}/\tau + \littleO{\sqrt{n}/\tau}}$ many queries.
However, we have
$2^{(\alpha+\littleO{1})\sqrt{n}/\tau} =
2^{(\alpha+\littleO{1}) m\cdot \frac{\sqrt{n}}{\tau m}} < 2^{C m}
$,
where the inequality comes from observing that $\frac{\sqrt{n}}{\tau m} 
=  \frac{50}{K}$ (so that it suffices to pick $\alpha$ satisfying $50\alpha/K 
< C$).  This contradicts 
\autoref{fact:bt:hardness} and proves the theorem.
\end{proof}

\begin{remark}[Improving this result]
Proposition 1 of \cite{MO03} gives a lower bound on the best
noise stability that can be achieved by any monotone function.
If this lower bound were in fact tight --- 
that is, there exists a family of monotone functions $\{f_r\}$ 
such that for all $\gamma\in[-1,1]$, 
$\nstab_{1-\gamma}(f_r) = (1-\gamma)^{(\sqrt{2/\pi} + \littleO{1})\sqrt{r}}$
--- then the above lower bound could be extended to an (almost) 
optimal range of $\tau$, i.e. $\tau\in[\Phi(n)/\sqrt{n},{{\frac 1 2} - c}]$
for $\Phi$ any fixed superconstant function.
\end{remark}

\medskip

\noindent {\bf From hardness of learning monotone functions 
to hardness of learning $k$-alternating functions.}  
We now establish the hardness of learning $k$-alternating functions.
Hereafter we denote by $\mathcal{H} = \{g_r\otimes\mathcal{G}_m\}_{m,r}$ 
the class of ``hard'' monotone functions from 
\autoref{theorem:strong:learn:lb:monotone}. Since $g_r$ is balanced and 
every $f\in\mathcal{G}_m$ has bias zero, 
it is easy to see that $\mathcal{H}$ is a class of balanced functions.

We begin by recalling the following useful fact about the noise stability 
of functions that are close to $\parity{}$:

\begin{fact}[e.g., from the proof of Theorem 9 in~\cite{BT13}]\label{fact:nstab:approx:parity}  Let $r\geq 1$. If $f$ is a Boolean function on $r$ variables which 
$\eta$-approximates $\parity{r}$, then for all $\delta\in[0,1]$,
  \begin{equation}
    \nstab_{1-2\delta}(f)     \leq (1-2\eta)^2(1-2\delta)^r + 4\eta(1-\eta).
  \end{equation}
\end{fact}

We use the above fact to define a function that is tailored to our needs: 
that is, a $k$-alternating function that is very sensitive to noise
\emph{and is defined on roughly $k^2$ inputs}. 
Without the last condition, one could just use $\parity{k}$, but in our
context this would only let us obtain a $\sqrt{k}$ (rather than a $k$)
in the exponent of the lower bound, because of the loss in the reduction. 
To see why, observe that by using a combining function on $k$ variables 
instead of $k^2$, the number of variables of the combined function 
$g_k\otimes\mathcal{G}_m$ would be only $n=km$.  However, to get a 
contradiction with the hardness of monotone functions we shall 
need $k\sqrt{n}/\eps \ll \sqrt{m}/\tau$, where $\tau\approx\eps/k$,
as the hardness amplification lemma requires the error to scale 
down with the number of combined functions.

\begin{definition}\label{def:k:monotone:parity}
For any {odd}\footnotemark{} $r\geq k \geq 1$, let $\parity{k,r}^\prime$ be the symmetric Boolean function on $r$ inputs defined as follows: for all $x\in\{0,1\}^r$,
  \[
      \parity{k,r}^\prime(x) = \begin{cases}
        0 & \text{if }\abs{x} \leq \frac{r-k}{2} \\
        1 & \text{if }\abs{x} \geq{\frac{r+k}{2}} \\
        \parity{r}(x) & \text{otherwise.}
      \end{cases}
  \]
In particular, $\parity{k,r}^\prime$ is $k$-alternating, and agrees with 
$\parity{r}$ on the ${k+1}$ middle layers of the hypercube. 
By an additive Chernoff bound, one can show that $\parity{k,r}^\prime$ is $\eta$-close to $\parity{r}$, for $\eta=e^{-k^2/2r}$.
\end{definition}
\footnotetext{The above definition can be straightforwardly extended to $r\geq k\geq 1$ not necessarily odd, resulting in
a similar $k$-alternating perfectly balanced function $\parity{k,r}^\prime$ that agrees with $\parity{r}$ on $k+\bigO{1}$ middle layers of the cube and is $0$ below and $1$ above those layers. For the sake of simplicity we leave out the detailed description of the other cases.}

\begin{proof}[Proof of \autoref{theorem:strong:learn:lb:kmonotone}]
$\mathcal{H}^{(k)}_n$ will be defined as the class 
${\parity{k,r}^\prime\otimes\mathcal{H}_m}$ for some $r$ and $m$ such 
that $n=mr$ (see below).  
It is easy to check that functions in $\mathcal{H}^{(k)}_n$ are balanced
and $k$-alternating.
We show below that for $n$ sufficiently large, $2\leq k < n^{1/14}$ 
and $\eps \in [(1/300) (k^{14}/n)^{1/6},{{\frac 1 2} - c}]$, 
learning $\mathcal{H}^{(k)}_n$ to accuracy $1-\eps$ requires 
$2^{\bigOmega{k\sqrt{n}/\eps}}$ membership queries.

By contradiction, suppose we have an algorithm $A$ learning for all $n,k,\eps$ 
as above the class of $k$-alternating functions to accuracy $1-\eps$ 
using $T_A(n,k,\eps) < 2^{\beta\frac{k\sqrt{n}}{\eps}}$ membership queries, 
where $\beta > 0$ is a universal constant to be determined during the analysis.
We claim that this implies that for infinitely many values of $m$, one 
can learn ${\mathcal{H}_m}$ to some range of accuracies with a number of 
membership queries contradicting the lower bound of 
\autoref{theorem:strong:learn:lb:monotone}.

Fix any $n$ large enough, $k$ and $\eps$ as above (which in particular 
impose $k =\bigO{ n^{1/14} }$). The constraints we impose on $m$, $r$ and $\tau$ are the following:
\begin{align}
  mr = n; \quad
  \expbias{\tau}{\parity{k,r}^\prime} + \eps &\leq 1- \eps; \quad
  m = \omega_n(1); \quad 
  \tau \geq \frac{1}{m^{1/6}};  \label{eq:cilrb:step2:constraints:tau:lb} \\
  { \beta k \frac{\sqrt{n}}{\eps} } &< {\alpha\frac{\sqrt{m}}{\tau}},  
\label{eq:cilrb:step2:constraints:contradiction}
\end{align}
\ignore{
\begin{align}
  mr &= n \label{eq:cilrb:step2:constraints:mrN}\\
  \expbias{\tau}{\parity{k,r}^\prime} + \eps &\leq 1- \eps \label{eq:cilrb:step2:constraints:expbias}\\
  m &= \omega_n(1) \label{eq:cilrb:step2:constraints:m:inf}\\
  \tau &\geq \frac{1}{m^{1/6}}  \label{eq:cilrb:step2:constraints:tau:lb} \\
  { \beta k \frac{\sqrt{n}}{\eps} } &< {\alpha\frac{\sqrt{m}}{\tau}}  \label{eq:cilrb:step2:constraints:contradiction}
\end{align}
}
where the constraints in 
\eqref{eq:cilrb:step2:constraints:tau:lb} are for us to apply the 
previous theorems and lemmas, while 
\eqref{eq:cilrb:step2:constraints:contradiction} is needed to ultimately 
derive a contradiction.

One can show that by taking 
$r \eqdef \flr{ \frac{k^2}{2\ln 5 }} \geq 1$ and
$\tau \eqdef \frac{100\eps}{r},$
the second constraint of \eqref{eq:cilrb:step2:constraints:tau:lb} 
is satisfied,
as then $\nstab_{1-\tau}(\parity{k,r}^\prime) \leq 1-8\eps$ 
(for the derivation, see Appendix~\autoref{ap:claim-setting-rtau}). 
Then, with the first constraint of \eqref{eq:cilrb:step2:constraints:tau:lb},
we get (omitting for simplicity the floors)
  $m \eqdef \frac{n\tau}{100\eps} = (2\ln 5)\frac{n}{k^2}$,
so as long as $k = \littleO{ \sqrt{n} }$, 
the third constraint of \eqref{eq:cilrb:step2:constraints:tau:lb} is met as
well.  With these settings, the final constraint of
\eqref{eq:cilrb:step2:constraints:tau:lb} can be rewritten as
$
\eps \geq \frac{1}{100} \left( \frac{r^{7}}{n} \right)^{1/6} = \frac{1}{100(2\ln 5)^{7/6}} \left( \frac{k^{14}}{n} \right)^{1/6}.
$
As $(2\ln 5)^{7/6} > 3$, it is sufficient to have
$\eps\geq \frac{1}{300} \left( \frac{k^{14}}{n} \right)^{1/6},$
which holds because of the lower bound on $\eps$.

It only remains to check Constraint~\eqref{eq:cilrb:step2:constraints:contradiction} holds:
\begin{align*}
  k \frac{\sqrt{n}}{\eps} &= 100k \frac{\sqrt{n}}{\tau r} = 
100 \frac{k}{\sqrt{r}} \frac{\sqrt{m}}{\tau} 
\leq \left( 100\sqrt{ \frac{ 2 \ln 5}{1-2 \ln 5/k^2} } \right) 
\frac{\sqrt{m}}{\tau} 
\leq 300\sqrt{  2 \ln 5 }\cdot\frac{\sqrt{m}}{\tau},
\end{align*}
where the first inequality holds because as 
$\frac{1}{r}\leq \frac{1} {\frac{k^2}{2\ln 5} -1}$
and the second holds because $k \geq 2.$
So for the right choice of $\beta=\bigOmega{1}$, e.g. $\beta=\alpha/600$, $\beta k \frac{\sqrt{n}}{\eps} < \alpha\frac{\sqrt{m}}{\tau}$, and \eqref{eq:cilrb:step2:constraints:contradiction} is satisfied.\medskip

It now suffices to apply \autoref{theo:rocco:xor:lemma} to 
${\parity{k,r}^\prime\otimes\mathcal{H}_m}$, with parameters $\gamma=\tau$ 
and $\eps$, on algorithm $A$, which has accuracy 
$\operatorname{acc}(A) \geq 1-\tau \geq \expbias{\gamma}{\parity{k,r}^\prime}
+\epsilon$.  
Since the functions of $\mathcal{H}$ are unbiased, it follows that there 
exists an algorithm $B$ learning $\mathcal{H}_m$ to accuracy $1-\tau$, 
with $\tau > 1/2m^{1/6}$, making only
\[
T_B(m,\tau) = \bigO{ T_A(n, k, \eps )\poly(n, k, {1}/{\eps}) } = 2^{ \beta k \frac{\sqrt{n}}{\eps}(1+\littleO{1}) } < 2^{\alpha\frac{\sqrt{m}}{\tau}}
\] membership queries, 
which contradicts the lower bound of 
\autoref{theorem:strong:learn:lb:monotone}.
\end{proof}

\begin{remark}[On the relation between $\eps$ and $k$]\label{remark:lb:against:kmonotone:tradeoff:k:eps}
  The tradeoff in the ranges for $k$ and $\eps$ appear to be inherent to this approach. Namely, it comes essentially from Constraint~\eqref{eq:cilrb:step2:constraints:tau:lb}, itself deriving from the hypotheses of \autoref{theorem:strong:learn:lb:monotone}. However, even getting an optimal range in the latter would still require $\tau=\bigOmega{1/\sqrt{m}}$, which along with $r\approx k^2$ and $\tau\approx\eps/r$ impose $k=\bigO{n^{1/6}}$ and $\eps=\bigOmega{k^3/\sqrt{n}}$.
\end{remark}

\bibliographystyle{alpha}	
\bibliography{main-flat}	

\appendix
\section{Proofs}\label{appendix:proofs:lb}

\subsection{Proof of \autoref{claim:high:influence:alternations}.} \label{ap:proof-influence}

Suppose ${\Inf[f]}\geq\alpha n$ for some $\alpha \in (0,1]$: this means that at least an $\alpha$ fraction of all edges are bichromatic. Define the \emph{weight level $k$} (denoted $\mathcal{W}_k$) to be the set of all edges going from a vertex of Hamming weight $k$ to a vertex of Hamming weight $k+1$ (in particular, $\abs{\mathcal{W}_k}=(n-k)\binom{n}{k}$), and consider weight levels $n/2-a\sqrt{n},\dots,n/2+a\sqrt{n}{-1}$ (the ``middle levels'') for $a\eqdef\sqrt{(1/2)\ln(8/\alpha)}$. 
{(We suppose without loss of generality that $n/2 - a \sqrt{n}$ is a whole number.)}
Now, the fraction of all edges which do \emph{not} lie in these middle levels is at most
\[
    \frac{1}{n 2^{n-1}}\cdot 2\sum_{j={0}}^{\frac{n}{2}-a\sqrt{n}{-1}}\abs{\mathcal{W}_k} \leq \frac{2n}{n 2^{n-1}}\sum_{j={0}}^{\frac{n}{2}-a\sqrt{n}{-1}} \binom{n}{k}
    \leq \frac{4}{2^{n}}\sum_{j=1}^{\frac{n}{2}-a\sqrt{n}{-1}} \binom{n}{k} \leq 4e^{-2a^2}=\frac{\alpha}{2}.
\]
So no matter how many of these edges are bichromatic, it must still be the case that at least an $\alpha/2$ fraction of all edges in the ``middle levels'' are bichromatic.

Since the ratio
\[
    \frac{ \abs{ \mathcal{W}_{n/2} } }{ \abs{ \mathcal{W}_{n/2-a\sqrt{n}} } } = \frac{ \frac{n}{2}\binom{n}{n/2} }{ \left(\frac{n}{2}+a\sqrt{n}\right)\binom{n}{n/2-a\sqrt{n}} }
\]
converges monotonically from below (when $n$ goes to infinity) to $C\eqdef e^{2a^2}$, any two weight levels amongst the middle ones have roughly the same number of edges, up to a multiplicative factor $C$. Setting $p=\alpha/6C$ and $q=\alpha/6$, this implies that at least a $p$ fraction of the weight levels in the middle levels have at least a $q$ fraction of their edges being bichromatic. (Indeed, otherwise we would have, letting $b_k$ denote the number of bichromatic edges in weight layer $k$,
\begin{align*}
\frac{\alpha}{2}\cdot \underbrace{\sum_{k=\frac{n}{2}-a\sqrt{n}}^{\frac{n}{2}+a\sqrt{n}{-1}} \abs{\mathcal{W}_k}}_{\text{total}} &\leq \sum_{k=\frac{n}{2}-a\sqrt{n}}^{\frac{n}{2}+a\sqrt{n}{-1}} b_k
\leq
\sum_{\substack{k\in[\frac{n}{2}-a\sqrt{n}, \frac{n}{2}+a\sqrt{n}{-1}]\\ b_k > q\abs{\mathcal{W}_k}}} \abs{\mathcal{W}_k} +
\sum_{\substack{k\in[\frac{n}{2}a\sqrt{n}, \frac{n}{2}+a\sqrt{n}{-1}]\\ b_k \leq q\abs{\mathcal{W}_k}}} q\cdot\abs{\mathcal{W}_k} \\
&\leq
p\cdot 2a\sqrt{n}\cdot\abs{\mathcal{W}_{n/2}} + q\cdot \sum_{k=\frac{n}{2}-a\sqrt{n}}^{\frac{n}{2}+a\sqrt{n}{-1}} \abs{\mathcal{W}_k} 
\leq p\cdot C \cdot \sum_{k=\frac{n}{2}-a\sqrt{n}}^{\frac{n}{2}+a\sqrt{n}{-1}} \abs{\mathcal{W}_k} + q\cdot \sum_{k=\frac{n}{2}-a\sqrt{n}}^{\frac{n}{2}+a\sqrt{n}{-1}} \abs{\mathcal{W}_k}.
\end{align*}
So $\frac{\alpha}{2}\cdot \text{total} \leq p\cdot C \cdot \text{total} + q\cdot. \text{total}
$, which gives
$\frac{\alpha}{2}\leq \frac{\alpha}{6C}\cdot C + \frac{\alpha}{6} = \frac{\alpha}{3}$, a  contradiction.)

Let $S$ be this collection of at least $2a\sqrt{n}p$ weight levels (from the middle ones) that each have at least a $q$ fraction of edges being bichromatic, and write $p_i$ to denote the fraction of bichromatic edges in $\mathcal{W}_i$, so that for each $i\in S$ it holds that $p_i \geq q$. Consider a random chain from $0^n$ to $1^n$. The marginal distribution according to which an edge is drawn from any given fixed weight level $i$ is uniform on $\mathcal{W}_i$, so by linearity, the expected number of bichromatic edges in a random chain is at least $\sum_{i\in S} p_i \geq 2a\sqrt{n}pq=\bigOmega{\sqrt{n}}$, and hence some chain must have that many bichromatic edges.
\qed

\subsection{Derivation of \autoref{theo:odonnell:function} using Theorem~3 of \cite{MO03}.} \label{sec:MO-changes}

The original theorem is stated for $\tau=1$, with the upper bound 
being $1-\bigOmega{1}$. However, the proof of \cite{MO03} 
goes through for our purposes until the very end, where they 
set $\epsilon\eqdef\frac{1}{\sqrt{r}}$ and need to show that 
\[ e^{-2}\left( 1-(1-\epsilon+2\sqrt{\epsilon/r})^{\sqrt{r}} \right)=
\bigOmega{1}. \]
    
More precisely, the proof goes overall as follows: for some realization of 
the Talagrand function on $r$ variables $g_r$, we 
want (for some absolute constant $K$) that
      \[
         1-K\tau \geq \nstab_{ 1-\frac{\tau}{\sqrt{r}} }(g_r)          = 1- 2\probaOf{ g_r\circ N_{ 1-\frac{\tau}{\sqrt{r}} }(x)\neq g_r(x) }
.
      \]
      That is, one needs to show
      $
      \probaOf{ g_r\circ N_{ 1-\frac{\tau}{\sqrt{r}} }(x)\neq g_r(x) } \geq \frac{K}{2}\tau
      $; and in turn, it is sufficient to prove that for $g$ a random Talagrand function on $r$ variables,
      \[
        \E_g\!\left[\ \probaOf{ g\circ N_{ 1-\frac{\tau}{\sqrt{r}} }(x)\neq g(x) }\ \right] \geq \frac{K}{2}\tau.
      \]
This is where we slightly adapt the \cite{MO03} proof. Where they set a 
parameter $\epsilon$ to be equal to ${1}/{\sqrt{r}}$ and analyze 
$\E_g\!\left[ \probaOf{ g\circ N_{ 1-2\epsilon }(x)\neq g(x) } \right]$,
we set for our purposes $\epsilon\eqdef\frac{\tau}{2\sqrt{r}}$. The rest of the argument goes through until the very end, where it only remains to show that
    \begin{equation}\label{eq:adapt:theo3:MO}
      a e^{-2}\left( 1-(1-\epsilon+2\sqrt{\epsilon/r})^{\sqrt{r}} \right) \geq \frac{K}{2}\tau
    \end{equation}
    ($a$ being a small constant resulting from the various conditionings
in their proof), or equivalently, that 
{$(1-\epsilon+2\sqrt{\epsilon/r})^{\sqrt{r}} \leq 1 - \frac{e^{2}K}{2a}\tau$}.
    But the left-hand side can be rewritten as
    \begin{align*}
      (1-\epsilon+2\sqrt{\epsilon/r})^{\sqrt{r}} &= e^{ \sqrt{r}\ln(1-\epsilon+2\sqrt{\epsilon/r}) }= e^{ \sqrt{r}\ln(1-\tau/2\sqrt{r}+\sqrt{2\tau}/r^{3/4}) } \\
      &= e^{ \sqrt{r}\ln\left(1- {\frac{\tau}{2\sqrt{r}}\left(1-\frac{2\sqrt{2}}{\sqrt{r^{1/2}\tau}}\right)} \right) } \\
      &\leq e^{ -\sqrt{r}\cdot{\frac{\tau}{2\sqrt{r}}\left(1-\frac{2\sqrt{2}}{\sqrt{r^{1/2}\tau}}\right)} } \tag{as ${\frac{\tau}{2\sqrt{r}}\left(1-\frac{2\sqrt{2}}{\sqrt{r^{1/2}\tau}}\right)}<1$} \\
      &= e^{ -\frac{\tau}{2}\left(1-\frac{2\sqrt{2}}{\sqrt{r^{1/2}\tau}}\right) }  \leq e^{ -\frac{\tau}{2}(1-\frac{1}{\sqrt{2}}) } \tag{as $\tau > \frac{16}{\sqrt{r}}$}\\
      &\leq e^{ -\frac{\tau}{7} } \leq 1-\frac{\tau}{8} \leq 1-\frac{e^{2}K}{2a}\tau \tag{first as $\tau < 1$, then for a suitable choice of $K$}.
    \end{align*}\qed

\subsection{Proof of \autoref{fact:bt:hardness}.} 
\label{ap:proof-fact-bt-hardness}

We give the proof for $m$ even; by standard techniques, it extends easily to 
the odd case. For any $m \in 2\mathbb{N}$, define $\mathcal{C}_m$ as the class 
of functions $f$ generated as follows: let $R=\setOfSuchThat{ x\in\{0,1\}^m }
{ \abs{x} = m/2 }$, and partition $R$ in $|R|/2$ pairs of 
elements $(x^\ell, \bar{x}^\ell)$. For all $x\in\{0,1\}^m$,
\[
  f(x) = \begin{cases}
    0 &\text{ if } \abs{x} < m/2\\
    r_\ell &\text{ if } x\in R\text{ and }x=x^\ell\\
    1-r_\ell &\text{ if } x\in R\text{ and }x=\bar{x}^\ell\\
    1 &\text{ if } \abs{x} > m/2
  \end{cases}
\]
where the $\abs{R}/2$ bits $r_\ell$ are chosen independently and uniformly 
at random. Clearly, $f$ is balanced, and we have
\[
    \abs{R} = \binom{ m }{ m/2 } \operatorname*{\sim}_{m\to\infty} \sqrt{\frac{2}{\pi}}\cdot\frac{2^m}{\sqrt{m}} \eqdef \gamma 2^m.
\]
Suppose we have a learning algorithm $A$ for $\mathcal{C}_m$ making 
$q < 2^{Cm}$ membership queries. Fix \mbox{$0 < \alpha \leq 1$}, and $\eps=\alpha/\sqrt{m}$; to achieve error at most $\eps$ overall, $A$ must in particular achieve error at most $\frac{\eps}{\gamma} =\sqrt{\frac{\pi}{2}}\alpha$ on $R$. But after making $q$ queries, there are still at least $t=\gamma 2^{m}/2-2^{Cm} > 0.99\abs{R}$ points in $R$ (for $m$ big enough) $A$ has not queried, and hence with values chosen uniformly at random; on each of these points, $A$ is wrong with probability exactly half, and in particular
\begin{align*}
  \probaOf{\text{error } \leq \frac{\eps}{\gamma} } &< \probaOf{\text{error } \leq 2\alpha } = \probaOf{ \sum_{i=1}^t X_i \leq 2\alpha \abs{R} } \\
  &\leq \probaOf{ \sum_{i=1}^t X_i \leq \frac{200}{99}\alpha t } \\
  & \leq e^{-\frac{(1-\frac{400}{99}\alpha)^2 t}{2}} = \littleO{1}
\end{align*}
with an additive Chernoff bound. This means that with high probability over the choice of the target concept, $A$ will fail to learn it to accuracy $1-\eps$.\qed

\subsection{Derivation of the bound 
\texorpdfstring{$\nstab_{1-\tau}(\parity{k,r}^\prime) \leq 1-8\eps$}{on the stability of the approximated parity}.}
\label{ap:claim-setting-rtau}
By setting $r$ as {stated} we get that $r \leq k^2/\ln(1/\eps)$ and the distance between 
$\parity{k,r}^\prime$ and $\parity{r}$ becomes $\eta=e^{-k^2/2r} \leq 1/5$. 
Since we aim at having $\expbias{\tau}{\parity{k,r}^\prime} \leq 1-2\eps$, 
it is sufficient to have $\sqrt{\nstab_{1-\tau}(\parity{k,r}^\prime)} 
\leq 1-4\eps$; which would in turn be implied by $\nstab_{1-\tau}
(\parity{k,r}^\prime)\leq 1-8\eps$.

By \autoref{fact:nstab:approx:parity}, it is sufficient to show that 
$(1-2\eta)^2(1-\tau)^r + 4\eta(1-\eta) \leq 1-8\eps$; note that 
since $\eps < 1/100$ and by our choice of $\tau$,
\begin{align*}
  (1-2\eta)^2(1-\tau)^r + 4\eta(1-\eta) &\leq \frac{(1-2\eta)^2}{1+100\eps} 
  + 4\eta(1-\eta) \leq (1-2\eta)^2(1-50\eps) + 4\eta(1-\eta) \\
  &\leq (1-4\eta+4\eta^2)(1-50\eps) + 4\eta(1-\eta) \\
  &= 1-4\eta-50\eps +200\eta\eps +4\eta^2-200\eps\eta^2 + 4\eta-4\eta^2 \\
  &= 1-50\eps +200\eps\eta(1-\eta) \leq 1-50\eps +32\eps
  = 1-18\eps\\
  &\leq 1-8\eps.
\end{align*}\qed

\end{document}